\documentclass[10pt,a4paper]{article}
\usepackage[left=0.5in,right=0.5in,top=0.5in,bottom=0.5in]{geometry} % <-- updated!
\usepackage{amsmath,amssymb}
\usepackage{graphicx}
\usepackage{float}
\usepackage[labelfont=bf]{caption}
\usepackage{titlesec}
\usepackage[colorlinks=true,allcolors=blue]{hyperref}
\usepackage{enumitem}
\usepackage{authblk}
\usepackage{subcaption}
\usepackage{tikz}
\usepackage{multirow}
\usepackage{xcolor}
\usepackage{graphicx}
\usepackage{epstopdf} 
\usepackage{cite}
\titlespacing*{\section}{0pt}{1.4em}{0.8em}
\titlespacing*{\subsection}{0pt}{1.0em}{0.5em}
\titleformat{\section}{\bfseries\large}{\thesection.}{0.5em}{}
\titleformat{\subsection}{\bfseries\normalsize}{\thesubsection.}{0.4em}{}

\title{Revealing Strain and Disorder in Transition-Metal Dichalcogenides Using Hyperspectral Photoluminescence Imaging}
\author[1, 2]{Adam Alfrey}
\author[2]{Cole Tait}
\author[3]{Takashi Taniguchi}
\author[4]{Kenji Watanabe}
\author[1, 2, 5]{Steven T. Cundiff\thanks{cundiff@umich.edu}}
\affil[1]{Applied Physics Program, University of Michigan, Ann Arbor, MI 48109, USA}
\affil[2]{Department of Physics, University of Michigan, Ann Arbor, MI 48109, USA}
\affil[3]{Research Center for Materials Nanoarchitectonics, National Institute for Materials Science,  1-1 Namiki, Tsukuba 305-0044, Japan}
\affil[4]{Research Center for Electronic and Optical Materials, National Institute for Materials Science, 1-1 Namiki, Tsukuba 305-0044, Japan}
\affil[5]{Quantum Research Institute, University of Michigan, Ann Arbor, MI 48109, USA}
\date{\today}

\begin{document}
\twocolumn[
\maketitle

\begin{abstract}
Hyperspectral photoluminescence (HSPL) imaging provides spatially resolved spectral information for monolayer transition-metal dichalcogenides (TMDs), enabling the detection of subtle variations in excitonic features that are not accessible with conventional optical or photoluminescence intensity imaging. We employ HSPL to map the microscopic spatial distribution of strain and disorder in hBN-encapsulated MoSe$_2$ and WSe$_2$ samples. Quantitative extraction of exciton, trion, and biexciton energies and linewidths reveals strain gradients and localized deformations, such as wrinkles and ripples. The technique allows for characterization of regions with uniform optical properties and identification of areas affected by micro-scale disorder, which may be missed by optical microscopy. Measurements on samples with different device architectures and fabrication processes demonstrate the general utility of hyperspectral PL imaging for assessing spatial heterogeneity and optoelectronic quality in two-dimensional materials.
\vspace{0.5cm}
\end{abstract}
]
% -- Double columns start directly under abstract. --

Atomically-thin materials, such as transition-metal di\-chal\-cogenides (TMDs), are promising platforms for optoelectronics and quantum photonics, owing to their strong interaction with light and highly tunable properties \cite{Mak2010, Splendiani2010_NanoLett, Mak2016, Wang2018}. Their atomic thinness and tunable optoelectronic properties support a wide range of potential technologies, from ultrathin and flexible transistors to next-generation photodetectors, flexible/strainable single-photon emitters, and quantum photonics~\cite{Manzeli2017, Sun2022, Klein2019, Gant2019, Iff2019, Frisenda2018}. Recent advances in fabrication enable the creation of high-quality, large-area monolayers and heterostructures, paving the way for scalable integration and commercialization~\cite{Joseph2023,Ross2013,Li2023, Chowdhury2020}.

Characterizing, understanding and, controlling spatial inhomogeneities, particularly those arising from local strain, wrinkles, and disorder, is needed to achieve these goals, especially as sample quality and dimensions improve. Large-scale monolayer samples produced by exfoliation are inherently susceptible to  strain and associated band structure modifications due to differential thermal contraction, lattice and/or moire mismatch (in the case of heterolayers), and even the polymer-assisted dry-transfer process itself~\cite{Frisenda2017, Liang2017, Gant2019, HenriquezGuerra2023, Kumar2024, Mennel2018}. These effects can strongly influence device performance and are often spatially heterogeneous, so visualizing and quantifying them with spatial resolution is especially helpful to avoid problematic areas for demonstrations of novel physical properties and phenomena.

To address this need, we employ hyperspectral photoluminescence (HSPL) imaging to probe and map the microscopic spatial variation of strain and disorder in hexagonal boron nitride (hBN)-encapsulated monolayers of molybdenum diselenide (MoSe$_2$) and tungsten diselenide (WSe$_2$). Unlike conventional photoluminescence (PL) intensity imaging, which may miss subtle, but functionally important, features, HSPL imaging captures the full spectrum as the excitation beam scans across the sample. This approach yields energy and linewidth information of the material and its resonances, providing a deeper understanding of local electronic structure and disorder than intensity-only measurements allow~\cite{Ross2014, Jakubczyk2018, Lindlau2018, Mennel2018, Psilodimitrakopoulos2018, Jakubczyk2019, Negri2019}.

By spatially resolving the spectral landscape, hyperspectral imaging uncovers hidden information encoded in the energies and linewidths of local optical transitions. This information enables direct visualization of microscale effects such as wrinkles, disorder, and strain gradients; features often invisible to white-light far-field optical microscopy or PL intensity-only mapping~\cite{Mouchliadis2021, Rivera2021, Mahdikhanysarvejahany2022, Purz2022, Wang2022, LPurz2024, Geilen2025}.

\begin{figure*}
\centering
\includegraphics[width=\linewidth]{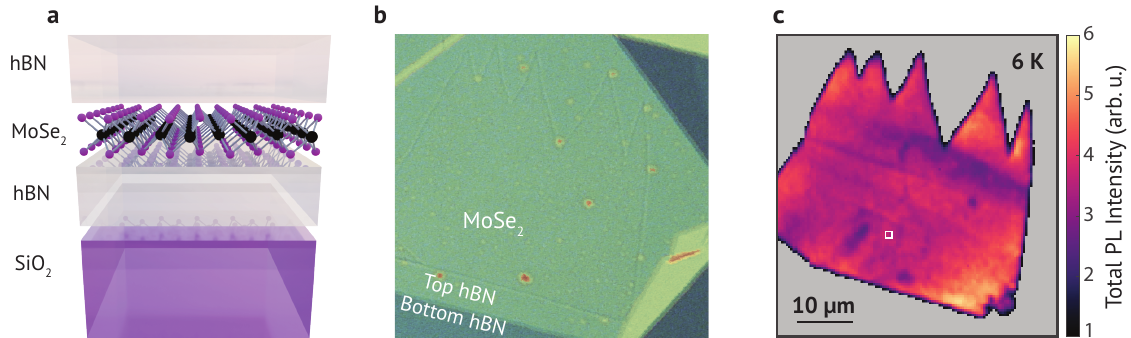}
\caption{\textbf{a}, Diagram of heterostructure labeled as Sample-1, consisting of an hBN encapsulated MoSe\textsubscript{2} monolayer. \textbf{b}, Microscope image of the heterostructure. \textbf{c}, Spatial map of the total PL intensity from the MoSe\textsubscript{2} at a temperature of 6 K. The spectrum at the location marked by the white box is shown in Fig.~\ref{fig:figure2}.}
\label{fig:figure1}
\end{figure*}

We first present PL on a large-area, residue-free hBN encapsulated MoSe$_2$ heterostructure, designated Sample-1, shown conceptually in Fig.~\ref{fig:figure1}a and via optical microscopy in Fig.~\ref{fig:figure1}b. See the methods section and supplemental material \cite{SupplementalMaterial} for further details on sample preparation. The raster-scanned total PL intensity, including both exciton and trion resonances, is shown in Fig.~\ref{fig:figure1}c. The flakes used in this sample were selected due to their isolated nature, and residue-free surfaces as indicated by the atomic-force microscopy measurements in Fig. X in the supplemental material \cite{SupplementalMaterial}.

\section*{Methods and materials}

Mechanically-exfoliated monolayer TMD crystals were first identified by conventional white light microscopy and then filtered by atomic force microscopy (AFM) results; selecting only the largest, most isolated, and residue-free flakes for device fabrication (see Supplemental information for AFM images \cite{SupplementalMaterial}. When clean flakes are verified, they are then stacked using a polymer-assisted dry-transfer technique and deposited onto SiO$_2$/Si substrates. The entire stacking process is carried out in a N$_{2}$ filled glovebox. This approach minimizes the trapping of air and water between the stack's layers, increasing sample cleanliness and flatness~\cite{Pizzocchero2016,Wang2013}.

Fig.~\ref{fig:figure1}c shows the total PL intensity at 6K using a cont\-inuous-wave (CW) HeNe laser at an excitation power of 20~$\mu$W, focused to a diffraction-limited spot with a Mitutoyo NA = 0.42 $50\times$ NIR objective. PL spectra were collected using a spectrometer equipped with a 600 grooves/mm grating and a thermo-electrically cooled (-80\textdegree C) CCD detector. Spatially resolved data sets were acquired in a raster fashion using a custom three-axis scanning stage equipped with Newport precision UMR12.40 stages and error-corrected LTA-HS linear actuators, which provide independent control of two steering mirrors (x and y) and the objective (z). Photoluminescence spectra were recorded at over ten thousand spatial locations across Sample-1, with measurements collected on a $62.5\mu\mathrm{m} \times 62.5\mu\mathrm{m}$ grid (141 steps per dimension, step size $\sim0.44~\mu\mathrm{m}$), ensuring sub-micron spatial resolution and complete sample coverage~\cite{Pawley2006}. Measurement on Sample-2 and Sample-3 were performed with the same apparatus using similar parameters including the spatial step size.

\section*{Quantitative Extraction of Resonance centers and widths}

\begin{figure}[H]
\centering
\includegraphics[width=0.75\columnwidth,height=0.65\columnwidth,keepaspectratio=true]{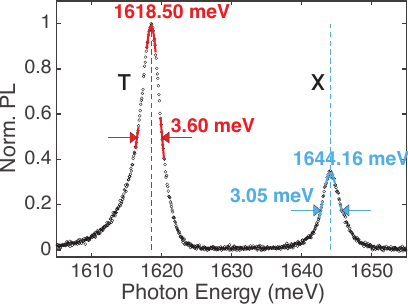}
\caption{\textbf{Quantitative parameter extraction.} Spectrum from a single location marked by white box in Fig.~\ref{fig:figure1}c, with trion (red) and exciton (blue) polynomial fits.}
\label{fig:figure2}
\end{figure}

A typical PL spectrum reveals two resonances, as shown in Fig.~\ref{fig:figure2} for Sample-1. The higher-energy peak is assigned to excitons, while the lower-energy peak is attributed to trions, excitons bound to an additional carrier. The widths of the resonances are due to a combination of homogeneous and inhomogeneous broadening. The homogeneous broadening arises from the finite lifetime of the excitations and elastic scattering by other excitations, including carriers and phonons. Inhomogeneous broadening arises from disorder, defects, and random strain due to wrinkles or nanobubbles. 

To construct spatially resolved maps, we fit the exciton and trion resonances at each pixel in the hyperspectral dataset. By fitting, we can obtain better estimates of the peak energy and width than simply taking the highest data point and data points closest to half the highest value. However, we do not fit the entire line as that requires assuming a lineshape function, which could bias the results, especially for asymmetric lineshapes such as that for the trion in Fig.~\ref{fig:figure2}. Rather, we use polynomials to fit localized regions around the peak and half maxima. Quadratics were fit to 16 data points around each peak maximum; the vertex of this polynomial was taken as the local resonance energy. The full-width at half-maxima (FWHM) of the resonances were evaluated by a linear fit to 8 points around the half maximum on each side, and using these to estimate the half-maximum values for determining the FWHM. Only spectra meeting a predefined signal-to-noise threshold were included in subsequent analysis and all other locations are set to grey pixels in the total PL intensity map. Likewise, if either of the polynomial fits failed, that pixel is set to grey in the corresponding center energy or FWHM map. 

This approach enabled robust, quantitative visualization of shifts in the resonance peak energies and widths across the sample, revealing the effects of local strain and microscale disorder with high spatial resolution. As we will show, mapping these spectral parameters reveals microscopic details that are not captured in a PL intensity map, which primarily reports macroscopic variations.

\section*{Results and discussion}

\begin{figure*}
	\centering
	\includegraphics[width=\linewidth]{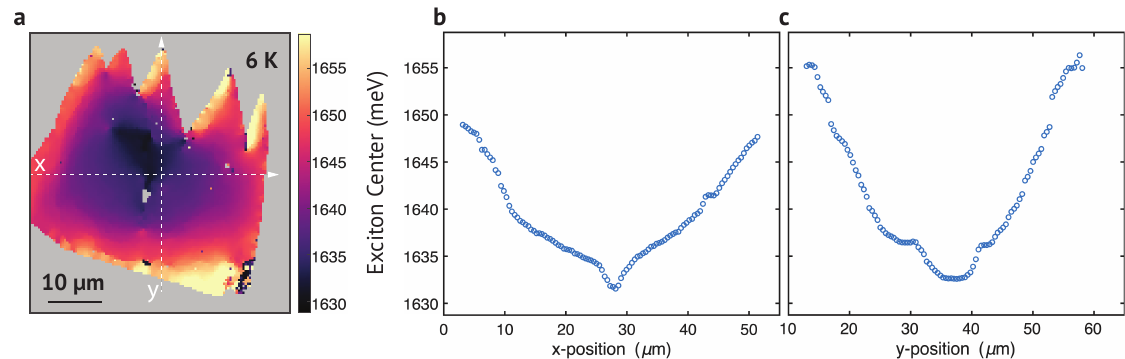}
	\caption{\textbf{a}, Spatial map of the center energy of the exciton resonance with dashed lines indicating the linecuts shown in \textbf{b}, Horizontal linecut. \textbf{c}, Vertical linecut.}
	\label{fig:figure3}
\end{figure*}

Using this approach, we generate an HSPL image of the exciton center energy for Sample-1. The HSPL image reveals a smooth, spatially varying redshift from the edges of the MoSe\textsubscript{2} toward the center (Fig.~\ref{fig:figure3}a), a behavior consistent with strain accumulation. This phenomenon can be attributed to differential thermal contraction: as the sample is cooled to cryogenic temperatures, the disparity in thermal expansion coefficients between the TMD monolayer and its underlying substrate (typically SiO$_2$/Si) yields a strain field across the stack~\cite{Liang2017, Gant2019, HenriquezGuerra2023}. Since, at room temperature, the thermal coefficient of expansion for MoSe\textsubscript{2} is 2-3 times greater than that of the silicon substrate, a free layer of MoSe\textsubscript{2} would shrink more than one attached to silicon as the temperature is reduced from room temperature to 6 K. Thus, we expect that the MoSe\textsubscript{2} is under radial tensile strain, similar to a drumhead.

Line scans of the exciton center energy (Fig.~\ref{fig:figure3}b, c) further reinforce the presence of a continuous strain gradient, showing a monotonic redshift from the periphery toward the central region. At the center, where the monolayer is more securely pinned because the atoms have nowhere to move, the tensile strain is maximal, locally narrowing the bandgap and producing a distinct redshift in the photoluminescence energy. This correlation between strain and modulation of the optical transition energy is robustly documented in TMDs and related van der Waals materials\cite{Liang2017, Gant2019, HenriquezGuerra2023, Kumar2024}.

\begin{figure}
	\centering
	\includegraphics[width=\columnwidth]{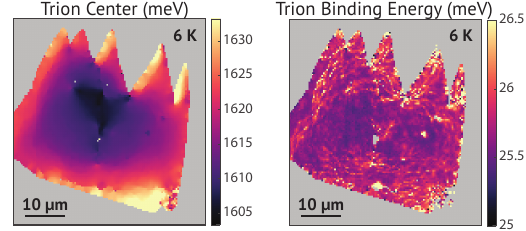}
	\caption{\textbf{Left:} Spatial map of the Trion fit center energy. \textbf{Right:} Map of Trion binding energy, highlighting the contrast between homogeneous regions and regions with wrinkles/ripples indicative of local strain and disorder.
	}
	\label{fig:figure4}
\end{figure}

The spatial map of the trion center energy reveals a similar redshift gradient, underscoring the sensitivity of both neutral and charged excitonic states to local strain (Fig.~\ref{fig:figure4} left). To distinguish the influence of strain from possible contributions due to extrinsic disorder or doping, we plot the trion binding energy, defined as the difference between the exciton center energy and the trion center energy, in the right panel of Fig.~\ref{fig:figure4}. Remarkably, the binding energy remains highly uniform throughout most of the sample (see supplemental information for a histogram of the binding eneries \cite{SupplementalMaterial}), highlighting the intrinsic stability of exciton-trion energetics in high-quality heterostructures. These results are consistent with a previous study that mapped the trion binding energy \cite{Wang2022}, although we observe lower levels of fluctuations in the binding energy.

Localized deviations in binding energy become apparent only in regions with pronounced deviations in the local strain, such as at wrinkles and ripples. These mechanical inhomogeneities, which are invisible in optical images or standard PL intensity maps, are revealed in the binding energy map, providing an effective marker for subtle but functionally important defects within the two-dimensional landscape.

\begin{figure}
	\centering
	\includegraphics[width=\columnwidth]{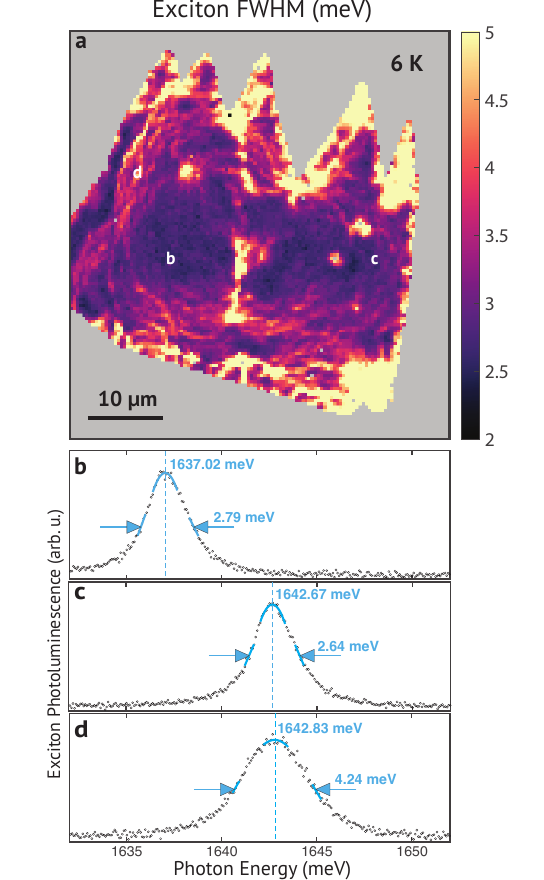}
	\caption{\textbf{a}, Exciton FWHM map with white letters indicating the locations of the spectra in panels b–d. \textbf{b}, Exciton spectrum from a central region with a red-shifted resonance. \textbf{c}, a blue-shifted edge region. \textbf{d}, a location with a pronounced wrinkle.}
	\label{fig:figure5}
\end{figure}

In contrast to the gradual shift in resonance energy, the spatial map of the exciton FWHM, shown in Fig.~\ref{fig:figure5}a, reveals a richer landscape of local variations. Across most of the sample, the FWHM remains relatively uniform, indicating homogeneous material with high quality and limited extrinsic disorder. However, distinct, spatially-localized increases in both the FWHM and the extracted binding energy emerge in select regions, as shown in Fig.~\ref{fig:figure5}a. These features are assigned to microscale wrinkles and ripples in the heterostructure, which are not visible in standard optical or PL intensity images but are evident here due to their influence on local strain and the electronic environment. These results show that the linewidth is more sensitive to the microscopic fluctuations than the center energy, which provides information on macroscopic variations in the strain landscape.

The left boundary of Sample-1 in the HSPL maps at 6K has shifted relative to the room-temperature optical micrograph Fig.~\ref{fig:figure1}b (see the Supplemental Material for an overlay of the two images \cite{SupplementalMaterial}). This microscale, cryogenically induced reconfiguration likely drove the formation of the undulating ripples observable in Fig.~\ref{fig:figure4} (right) and Fig.~\ref{fig:figure5}a. 

These localized spectroscopic signatures of strain serve as sensitive fingerprints for wrinkles and defects. Notably, the spectroscopic integrity of the exciton resonance remains intact across the sample, as demonstrated by the unchanging lineshapes between the heterostructure center (Fig.~\ref{fig:figure5}b) and the edge (Fig.~\ref{fig:figure5}c). This invariance demonstrates that intrinsic optical quality is preserved even in strained regions, supporting reliable comparison of energetic trends. The visualization of consistent lineshape regions also enables precise micro-positioning of the excitation beam for future measurements, including accurate alignment of multiple excitation sources. For comparison, Fig.~\ref{fig:figure5}d shows the exciton spectrum at a pronounced wrinkle, which causes a 40\% increase in the linewidth.

\begin{figure*}
\centering
\includegraphics[width=\linewidth]{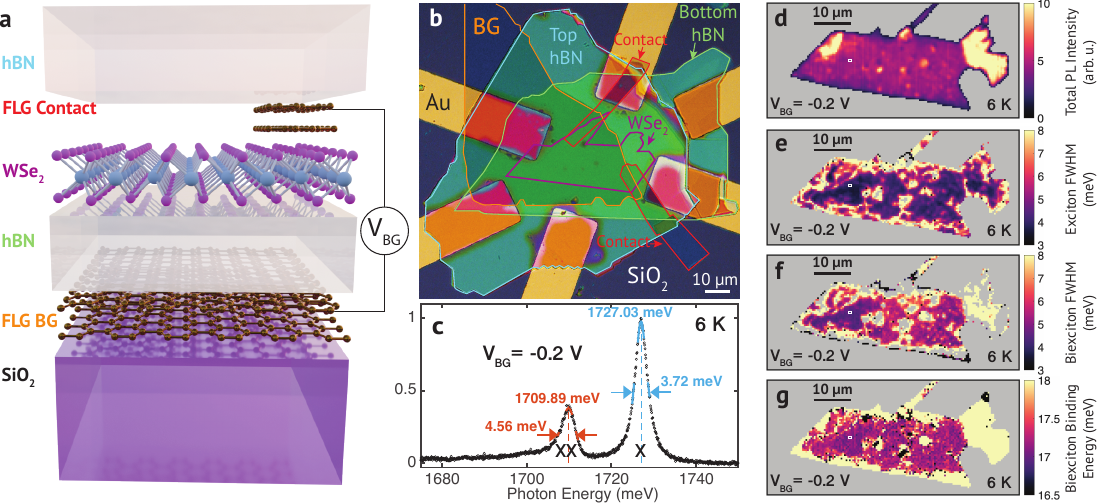}
\caption{\textbf{a}, Diagram of electrostatically gated WSe$_2$ heterostructure, designated as Sample-2 \textbf{b}, Microscope image of the heterostructure on the substrate containing pre-defined electrodes. \textbf{c}, Intrinsic PL spectrum from the location marked by the white boxes in panels d-g showing the neutral exciton (blue) and biexciton (orange) resonances. \textbf{d}, Total PL intensity of the intrinsically doped WSe$_2$ monolayer, which contains both biexciton and exciton integrated resonances. \textbf{e}, Exciton FWHM. \textbf{f}, Biexciton FWHM. \textbf{g}, Biexciton binding energy.}
\label{fig:figure6}
\end{figure*}

To further demonstrate HSPL imaging on TMDs, we examined a monolayer WSe$_2$ heterostructure encapsulated in hBN, denoted as Sample-2,  with electrical contacts provided via few-layer graphite (FLG) electrode contacts and an FLG back gate (BG) (Figs. \ref{fig:figure6}a, b). The contacts are about 5 layers of graphene with the back gate being 10-20 layers thick. While the device contains several functional layers, it is helpful to focus on the role of the bottom hBN, which acts as the dielectric in a double-plate capacitor, specifically separating the contacts and WSe$_2$ from the back gate. This architecture allows for continuous electrostatic gating and precise control over carrier density, enabling \textit{in situ} exploration of p-type ($V_{BG}\lessapprox-1 V$), intrinsic ($V_{BG}\approx-0.2 V$), and n-type ($V_{BG}\gtrapprox0.5 V$) doping regimes. To set the WSe$_2$ monolayer as close to intrinsic as possible, the back gate voltage was fixed at $V=-0.2 V$ for the spectrum shown in Fig.~\ref{fig:figure6}c and the spatial maps presented in Figs.~\ref{fig:figure6}(d-g).

PL spectroscopy as a function of gate voltage (See Supplemental Information \cite{SupplementalMaterial}) reveals the emergence of distinct optical resonances due to various excitations in each regime: negative singlet ($X_{S}^-$) and triplet ($X_{T}^-$) trions in n-doped regions, neutral excitons ($X^0$) and biexcitons ($XX$) in the intrinsic regime (Fig.~\ref{fig:figure6}c), and positive trions ($X^+$) under p-type doping~\cite{Ye2018, Barbone2018, Li2018}.

A typical PL spectrum for Sample-2 at $V_{BG}\approx-0.2 V$, corresponding to intrinsic material with no electrons or holes, is shown in Fig.~\ref{fig:figure6}c. Similar to Sample-1, two resonances are observed. Again, the high-energy peak is assigned to excitons, whereas the low-energy peak is assigned to biexcitons, bound molecular states of two excitons, rather than trions, as no carriers are present to bind to the excitons.

It is interesting to note that the backgate does not extend beneath the ``tail'' on the right-hand side of the WSe$_2$ monolayer, which is hyperspectrally visible in Fig.~\ref{fig:figure6}(d-g). In this region, the carrier concentration remains uncontrolled by the gate and thus exhibits n-type behavior, which is the natural state of this particular crystal in the absence of gating, as evidenced by the brighter total PL intensity. These regions uncontrolled by the gate should be ignored in the biexciton maps Fig.~\ref{fig:figure6}(f, g).

We present intrinsically-doped spatial maps of PL intensity, FWHMs of the WSe$_2$ exciton and biexciton, and the biexciton binding energy in Fig.~\ref{fig:figure6} (d-g). Our results show that, as with the MoSe$_2$ samples, HSPL imaging captures both smooth band structure modulations and microscopic disorder features even with more elusive quantum states such as the biexciton. The biexciton maps presented here exemplify how hyperspectral PL imaging can provide confidence in identifying the presence of this four-particle state and determining which regions have signals robust enough to warrant further investigation. The biexciton and positive trion binding energy frequency histograms (see supplemental information \cite{SupplementalMaterial}) also show that HSPL can be used to determine the binding energy of the quasiparticles present in the system in a statistically meaningful manner.

A third sample, Sample-3, consists of hBN-encapsulated MoSe$_2$ that was nano-squeegeed to remove residue and bubbles remaining after the transfer process (see supplemental information \cite{SupplementalMaterial}). Although the sample appears clean under optical inspection, HSPL maps reveal a different story: micro-wrinkles and ripples are widespread in the FWHM map, and the trion binding energy varies significantly across the sample. Sample-3 is detailed in the Supplementary Information for brevity, but demonstrates that HSPL imaging remains effective even when resonances are homogeneously broadened.

These observations further underscore the broad applicability of HSPL imaging for analyzing optoelectronic landscapes in van der Waals materials, including those processed by different fabrication techniques. The inclusion of this additonal sample highlights the robustness of hyperspectral PL methods in identifying subtle disorder and strain that conventional microscopy may miss.
\vspace{0.5cm}

In summary, hyperspectral photoluminescence imaging provides a versatile approach for mapping strain, disorder, and complex excitonic phenomena in monolayer TMDs with spatial and spectral resolution. By extracting exciton, trion, and biexciton energies and linewidths from different hBN-encapsulated TMD monolayers and device architectures, we identify both macroscopic smooth strain profiles and microscopic localized defects such as wrinkles and ripples, features typically invisible to conventional microscopy.

The ability to spatially resolve biexciton states and quantify their binding energies demonstrates that HSPL imaging can probe and characterize even subtle many-body quantum states. Our results show that this method remains effective for samples with non-pristine resonances, offering robust insights into material quality and device functionality across diverse fabrication techniques.

Extending these measurements to electrically gated WSe$_2$ heterostructures further highlights the utility of HSPL in revealing spatial variations of electronic properties, including carrier density modulations and the formation of engineered p-n junctions. The addition of a control sample subjected to nano-squeegeeing underscores the technique’s sensitivity to micro-scale disorder.

As van der Waals heterostructures and quantum materials research continue to evolve, non-invasive, hyperspectral PL imaging will play an increasingly vital role in advancing fundamental understanding and enabling the development of next-generation optoelectronic and quantum devices.

\section*{Acknowledgments}
The work at the University of Michigan work was supported by the US Department of Energy Grant DE-SC0022179. K.W. and T.T. acknowledge support from the JSPS KAKENHI (Grant Numbers 21H05233 and 23H02052) , the CREST (JPMJCR24A5), JST and World Premier International Research Center Initiative (WPI), MEXT, Japan.

\section*{Data availability}
The data generated in this study are available from the corresponding author upon reasonable request.

\bibliography{fb}
\bibliographystyle{titles}

\end{document}